# GENFIRE: A generalized Fourier iterative reconstruction algorithm for high-resolution 3D imaging

Alan Pryor, Jr.[1*], Yongsoo Yang[1*], Arjun Rana[1], Marcus Gallagher-Jones[1,2], Jihan Zhou[1], Yuan Hung Lo[1,3], Georgian Melinte[1,4], Wah Chiu[5], Jose A. Rodriguez[2] and Jianwei Miao[1]

[1]*Department of Physics and Astronomy and California NanoSystems Institute, University of California Los Angeles, California 90095, USA.* [2]*Department of Chemistry & Biochemistry, UCLA-DOE Institute of Genomics and Proteomics, Los Angeles, California 90095-1570, USA.*
[3]*Department of Bioengineering, University of California, Los Angeles, CA 90095, USA.*
[4]*Institut de Physique et Chimie des Matériaux de Strasbourg, CNRS-Université de Strasbourg, 23, rue du Loess, 67037 Cedex 08 Strasbourg, France.* [5]*Program in Structural and Computational Biology and Molecular Biophysics, Baylor College of Medicine, Houston, Texas 77030, USA.*
[*] *These authors contributed equally to this work.*
*Correspondence and requests for materials should be addressed to J.M. (email: miao@physics.ucla.edu)*

**Tomography has made a radical impact on diverse fields ranging from the study of 3D atomic arrangements in matter to the study of human health in medicine. Despite its very diverse applications, the core of tomography remains the same, that is, a mathematical method must be implemented to reconstruct the 3D structure of an object from a number of 2D projections. In many scientific applications, however, the number of projections that can be measured is limited due to geometric constraints, tolerable radiation dose and/or acquisition speed. Thus it becomes an important problem to obtain the best-possible reconstruction from a limited number of projections. Here, we present the mathematical implementation of a tomographic algorithm, termed GENeralized Fourier Iterative REconstruction (GENFIRE). By iterating between real and reciprocal space, GENFIRE searches for a global solution that is concurrently consistent with the measured data and general physical constraints. The algorithm requires minimal human intervention and also incorporates angular refinement to reduce the tilt angle error. We demonstrate that GENFIRE can produce superior results relative to several other popular tomographic reconstruction techniques by numerical simulations, and by experimentally by reconstructing the 3D structure of a porous material and a frozen-hydrated marine cyanobacterium. Equipped with a graphical user interface, GENFIRE is freely available from our website and is expected to find broad applications across different disciplines.**

Tomography has found widespread applications in the physical, biological and medical sciences[1–7]. Electron tomography, for example, is experiencing a revolution in high-resolution 3D imaging of physical and biological samples. In the physical sciences, atomic electron tomography (AET) has been developed to determine the 3D atomic structure of crystal defects such as grain boundaries, anti-phase boundaries, stacking faults, dislocations, chemical order/disorder and point defects, and to precisely localize the 3D coordinates of individual atoms in materials without assuming crystallinity[1,8–12]. The atomic coordinates measured by AET have been used as direct input to density functional theory calculations to correlate crystal defects and chemical order/disorder with material properties at the single atomic level[13]. In the biological sciences, single-particle cryo-electron microscopy (EM) has been applied to achieve near atomic resolution of purified protein complexes[2,7,14–16], and cryo-electron tomography allows for 3D imaging of pleomorphic samples such as viral infection mechanisms of cells with resolutions on the order of a few nanometers[17–19]. These advances are not limited to electron tomography. Tomographic implementation of synchrotron X-ray absorption and phase contrast imaging has also found interdisciplinary applications[5,20–25]. Using the brilliance of advanced X-ray sources, coherent diffractive imaging (CDI) methods[26] have been combined with tomographic reconstruction for 3D quantitative imaging of thick samples with resolutions in the tens of nanometers[27–33].

Presently, a popular tomographic reconstruction method is filtered back projection (FBP)[2–4]. FBP works well when there are a large number of projections with no missing data. However, when the data is inadequately sampled due to the radiation dose and geometric constraints, it suffers from artifacts, potentially clouding interpretability of the final reconstruction. This difficulty can be partially alleviated by real-space iterative algorithms such as the algebraic reconstruction technique (ART)[34], simultaneous ART (SART)[35] and simultaneous iterative reconstruction technique (SIRT)[36]. However, these algorithms do not fully exploit the correlated information among all the projections as the iteration process is implemented through local interpolation in real space. In contrast, Fourier-based iterative algorithms use information in both real and Fourier space as part of the iterative process[13,37,38]. A major advantage of these algorithms is that changes made in one space affect the other space globally. Equal slope tomography (EST)[37], an example of such an algorithm, has been successfully applied in AET to reconstruct the 3D arrangement of crystal defects in materials, including recovery of Bragg peaks in the missing wedge direction[1,8–10]. Additionally, EST was shown to produce reconstructions comparable to modern medical CT techniques but using significantly lower radiation dose[20,22,39]. However, the drawback of EST is the requirement that the tilt angles must follow equal slope increments along a single tilt axis, which limits its broader applications.

Very recently, a generalized Fourier iterative reconstruction algorithm (GENFIRE) has been reported for high-resolution 3D imaging with a limited number of 2D projections[13]. GENFIRE first pads zeros to each 2D projection and calculates its oversampled Fourier slice[40,41]. The oversampled Fourier slices are used to accurately compute a small fraction of points on a 3D Cartesian grid based on gridding interpolation[42,43]. The remaining grid points that cannot be determined with sufficient accuracy are defined as unknown. The algorithm then iterates between real and reciprocal space and enforces constraints in each space. In real space, the negative valued voxels and the voxels in the zero-padding region are set to zero. In reciprocal space, the small fraction of the known grid points are enforced in each iteration, while the unknown grid points are recovered by the iterative process. After several hundred iterations, the algorithm converges to a structure that is concurrently consistent with the measured data and the physical constraints. Furthermore, GENFIRE implements an angular refinement routine to reduce the tilt angle error and can be adapted to any tomographic data acquisition geometry. In this article, we present the mathematical implementation of the GENFIRE algorithm. Using both physical and biological samples, we demonstrate that GENFIRE produces superior 3D reconstructions relative to several other tomographic reconstruction algorithms.

## Methods

**Assembling a 3D Fourier grid with oversampling.** GENFIRE first assembles a rectangular 3D Fourier grid from a set of measured 2D projections. According to the Fourier slice theorem, the Fourier transform of each 2D projection represents a plane slicing through the origin of the 3D

Fourier transform of the sample. To obtain a Fourier grid point, $F_{obs}(\vec{k})$, we compute its perpendicular distance to the Fourier plane, where $D_j$ represents the perpendicular distance and $(u_j, v_j)$ the foot of the perpendicular line to the $j^{th}$ projection. Since $(u_j, v_j)$ are not integer coordinates, we use the discrete Fourier transform (DFT) to compute the value of $(u_j, v_j)$. The use of the DFT to explicitly compute the $(u_j, v_j)$ value is more accurate than interpolating from the set of 2D FFTs of each projection at the cost of being computationally slower. A faster, but less accurate, FFT gridding method is also provided as an option in the GENFIRE package. After calculating the $(u_j, v_j)$ values for all the projections, we determine the value of the Fourier grid point by

$$F_{obs}(\vec{k}) = \sum_{\{j\,|\,D_j < D_{th}\}} \frac{D_j^{-1}}{\sum_{\{j\,|\,D_j < D_{th}\}} D_j^{-1}} \sum_{x=-\frac{N}{2}}^{\frac{N}{2}-1} \sum_{y=-\frac{N}{2}}^{\frac{N}{2}-1} f_{obs}^j(x,y) e^{\frac{-2\pi i(xu_j+yv_j)}{NO}} \quad (1)$$

where $D_{th}$ is a predefined threshold, $f_{obs}^j(x,y)$ is the $j^{th}$ 2D projection with a size of N x N pixels and O is the linear oversmapling ratio[40,41]. By properly choosing $D_{th}$ and O, we accurately determine a small fraction of the Fourier grid points, while the remaining grid points are defined as unknown. In the GENFIRE package, the default value for $D_{th}$ and O is 0.5 and 3, respectively.

**The Fourier based iterative algorithm.** Due to radiation dose and/or geometric constraints, it is desirable in many tomography applications to achieve high-resolution 3D imaging from a limited number of projections. As a result, a significant amount of the assembled Fourier grid points remain unknown after the gridding process. To recover the unknown grid points, GENFIRE iterates between real and reciprocal space with general constraints enforced in each space (Fig. 1). For the 1st iteration, the values of the unknown grid points can be assigned to zero, random numbers or some other pre-determined numbers as the algorithm is not very sensitive to the initial input. The $j^{th}$ GENFIRE iteration consists of following five steps (Fig. 1).

(i) Apply the inverse FFT to $F_j(\vec{k})$ and obtain the $j^{th}$ image, $\rho_j(\vec{r})$.
(ii) Modify the image by applying the following constraints,

$$\rho_j'(\vec{r}) = \begin{cases} 0 & (\vec{r} \notin S) \cup (\rho_j(\vec{r}) < 0) \\ \rho_j(\vec{r}) & Otherwise \end{cases} \quad (2)$$

Where S represents a support, separating the zero-padding region from the sample structure. The zero-padding region is due to oversampling[40,41]. This step sets the voxels outside the support or negative valued voxels inside the support to zero, while retaining the values of the other voxels.

(iii) Apply the FFT to $\rho_j'(\vec{r})$ to obtain $F_j'(\vec{k})$.
(iv) Compute $F_{j+1}(\vec{k})$ by enforcing the Fourier space constraint,

$$F_{j+1}(\vec{k}) = \begin{cases} F_{obs}(\vec{k}) & Known\ voxels \\ F_j'(\vec{k}) & Unknown\ voxels \end{cases} \quad (3)$$

$F_{j+1}(\vec{k})$ is used for the $(j+1)^{th}$ iteration.
(v) Calculate two R-factors, $R_k$ and $R_{free}$,

$$R_k = \frac{\sum_{\vec{k}_{known}} |F_{obs}(\vec{k}) - F_j(\vec{k})|}{\sum_{\vec{k}_{known}} |F_{obs}(\vec{k})|} \quad (4)$$

$$R_{free} = \frac{\sum_{\vec{k}_{withheld}} |F_{obs}(\vec{k}) - F_j(\vec{k})|}{\sum_{\vec{k}_{withheld}} |F_{obs}(\vec{k})|} \quad (5)$$

where $\vec{k}_{known}$ represents the known voxels and $\vec{k}_{withheld}$ is a small number of randomly selected known voxels that are not used in the reconstruction. $R_k$ is an error metric to monitor the convergence of the iterative process, while $R_{free}$ is an unbiased free parameter to evaluate the reconstruction, which is used in crystallography[44]. $R_{free}$ is always larger than $R_k$, but for a good reconstruction the two R-factors should be consistent. Significant deviation could indicate overfitting. The algorithm is reliable and usually converges within several hundred iterations.

In the GENFIRE package, there is also an option to use resolution extension/suppression. For experimental data, the signal to noise ratio decreases with the increase of the spatial frequency. To compensate the high noise level at the high spatial frequency, we implement a resolution extension/suppression technique capable of partially decoupling signal and noise through a simple modification of the way the Fourier constraint is applied. For the first iteration, only the lowest spatial frequency information is enforced. As iterations progress, higher spatial frequency data is gradually applied. This continues, forming the extension step, until half of the total number of iterations has been completed, at which point all measured data is enforced. The process is then reversed for the second half of the reconstruction, and the spatial resolution of the enforced data is gradually reduced to form the suppression step until the final iteration when only the lowest frequency information is constrained once again. While resolution extension has been implemented before[45,46], to our knowledge, resolution extension/suppression has not been previously reported. We have performed extensive numerical simulations and observed that this technique can consistently improve the 3D reconstruction with noisy data (Supplementary Fig. 1). Full exploration of the generality of resolution extension/suppression requires follow-up studies.

**Angular refinement.** The experimentally measured tilt angles may not always coincide with the true orientations of the projections. This could be the result of many causes including instrument misalignment, slipping, beam-induced motion, vibration, thermal effects, or software error. To achieve high-resolution 3D reconstruction, we implement an angular refinement procedure to reduce the tilt angle error, which consists the following four steps.
(i) An initial 3D reconstruction is computed using the experimentally measured tilt angles.
(ii) For the $j^{th}$ projection, a series of 2D projections are calculated from the 3D reconstruction by varying the three Euler angles: $\phi \in [\phi_j - \delta\phi, \phi_j + \delta\phi]$, $\theta \in [\theta_j - \delta\theta, \theta_j + \delta\theta]$, $\psi \in [\psi_j - \delta\psi, \psi_j + \delta\psi]$, where $(\phi_j, \theta_j, \psi_j)$ are the current best fit for the Euler angles of the $j^{th}$ projection. Each calculated 2D projection is then compared with the corresponding measured projection, $f_{obs}^j(x, y)$, and a quality-of-fit metric is computed. The quality-of-fit metric can be implemented by either the normalized cross-correlation or the real space R-factor. For the latter, additional translational alignment between two projections have to be performed, whereas using cross correlation the translational search is performed simultaneously. The three Euler angles with either the largest cross correlation or smallest R-factor are recorded as the refined angles for the $j^{th}$ projection.

(iii) Repeat step (ii) for all the projections and a series of the refined angles are obtained.

(iv) Obtain a new 3D reconstruction with the refined angles for all the projections.

(v) Repeat steps (ii) – (iv) until no further improvement can be made.

In practice, each projection is refined in parallel, and the calculation of 2D projections from the 3D reconstruction represents the bulk of the computation. This calculation is expedited by applying the FFT to obtain an oversampled Fourier transform from the 3D reconstruction. Central slices are computed from the 3D Fourier transform using the C++ library splinterp for multithreaded linear interpolation. The inverse FFT is used to invert the central slices to the corresponding 2D projections. Care should be taken that while GENFIRE's reconstruction can find a global minimum, the current angular refinement approach may be trapped into local minima. Further developments are needed to search for a global minimum for angular refinement.

## Results

**Numerical simulations on the reconstruction of a biological vesicle.** Numerical simulations on the 3D reconstruction of a 64x64x64 voxel vesicle model (Figs. 2a-c) were performed using GENFIRE, EST, FBP and SIRT. Simulated projections were obtained by first calculating 2D Fourier slices of the 3D model for given angles. The corresponding real-space projections were then computed by applying the inverse FFT to the Fourier slices. This code is also included in the GENFIRE package and can be accessed graphically using the *Projection Calculator*. To evaluate the performance of various reconstruction algorithms with noise, we calculated 71 projections with the tilt angles ranging -70.1° to +70.1°. Noise was added to the projections at levels similar to that observed in cryo-EM images of cellular structures. Each set of projections were reconstructed using GENFIRE, EST, FBP and SIRT. The EST and GENFIRE reconstructions were performed using a loose support, the positivity constraint and 250 iterations. The SIRT reconstruction was achieved with the positivity constraint, long-object compensation and 125 iterations.

Figures 2d, g, j and m show a 10-voxel-thick central slice of the 3D reconstructions in the XY plane using GENFIRE, EST, FBP and SIRT, respectively, where the z-axis is the missing wedge direction. Because there is no missing data in this direction, the reconstructions from all methods exhibit good agreement with the model (Fig. 2a). However, along the missing wedge direction both GENFIRE and EST reconstructions (Figs. 2e, f, h and i) appear to be more isotropic and contain more fine features than FBP and SIRT (Figs. 2k, l, n and o). The Fourier shell correlation (FSC)[2] between the reconstructions and the model further confirms that the GENFIRE resconstruction is superior at all spatial frequencies compared to other algorithms. This simulation was also performed with no noise and higher noise (Supplementary Figs. 2 and 3). In the case of noise-free data with equal slope angles, EST produces slightly better results than GENFIRE as no interpolation is needed in EST. However, in practice this idealized scenario does not occur, and our results show that for even moderate noise levels GENFIRE produces better results. SIRT introduces a form of regularization to the reconstruction, which reduces missing wedge artifacts but also appears to compromise the resolution. By accurately assembling a small fraction of the Fourier grid points and using an iterative algorithm with resolution extension/suppression, GENFIRE is able to simultaneously reduce the effect of noise and retain higher resolution information. This capability will be important as scientists continue to solve important problems by pushing imaging systems to their limits.

**Numerical simulations on atomic electron tomography.** To quantify the GENFIRE reconstruction of 3D nanostructures at atomic resolution with noise and a missing wedge, we generated a 3D atomic model consisting of a 4.3 nm FePt$_3$ nanoparticle with a chemically ordered face-centered cubic (L1$_2$) phase. Using this model, 27 annular dark field (ADF) projections were computed using multislice simulation[47] (electron energy: 300 keV, probe size: 0.5 Å, C$_3$: 0 mm, C$_5$: 5 mm, probe convergence semi-angle: 30 mrad, and the inner and outer detector angles: 48 mrad and 251 mrad). The angular tilt range is ±70.1° and the pixel size is 0.4 Å. For each tilt angle, a total of 10 frozen phonon configurations were averaged. To simulate the convolution effect resulting from finite probe size and other incoherent effects, each image was convolved with a 2D Gaussian function with σ = 0.51Å. Poisson-Gaussian noise was then added to the ADF scanning transmission electron microscopy (STEM) projections.

After denoising was applied to the projections[48], this tilt series was reconstructed with GENFIRE, EST, SIRT and FBP, and the results are shown in Fig. 3. Visually, GENFIRE, EST, and SIRT all demonstrate reduction of reconstruction artifacts, though the difference appears more substantial for GENFIRE and EST (Figs. 3a-h). Both SIRT and FBP suffer from aliasing artifacts that produce what appear to be atoms, but are not actually present in the model, outside of the true boundary of the particle (Figs. 3c, d, g and h). These phantom atoms would prove problematic for atom tracing and refinement in AET. The iterative methods have also successfully recovered missing information as indicated by the presence of Bragg peaks in the missing wedge (magenta arrows in Figs. 3i-k). Determination of 3D atomic coordinates is most accurate when the reconstruction is isotropic, thus it is important for the reconstruction algorithm to be robust to noise and the missing wedge problem. Among the four algorithms, GENFIRE produces the best reconstruction of the 3D atomic structure.

**Angular refinement simulations.** To demonstrate the improvement made by angular refinement, a simulation was performed using the same 27 ADF-STEM projections from Fig. 3. The orientation angle of each projection was randomly shifted up to ± 2°, and a random translational shift of ± 1 pixel was applied along the x and y-axes. A preliminary GENFIRE reconstruction was performed and used as input to the refinement loop which was run for a total of 5 iterations with an angular search range of ± 3° with 0.2° steps, and with normalized cross-correlation as the error metric. The results of this simulation are shown in Fig. 4. The initial and refined angles were compared with the true ones using a normalized angular distance[49] (Fig. 4a.), resulting in an improvement from an initial average angular error of 2.1° to a refined value of 1.3°. The reconstruction is improved after angular refinement, shown in Figs. 4b and 4c. The boundary of the nanoparticle is also better defined, with fewer artifacts around the periphery.

**GENFIRE reconstruction on experimental data of a porous material.** To perform a quantitative comparison between GENFIRE and other iterative algorithms on experimental data, we acquired a tomographic tilt series of a Mo$_3$Si alloy annealed at 1100° C. Mo-Si and Mo-Si-B alloys are resistant to oxidation and creep and are among potential candidates with high melting temperatures to replace currently widely used Ni-based superalloys[50,51]. The experiment was conducted on an FEI TitanX 60-300 in STEM mode equipped with a Gatan high-angle annular dark field detector. The microscope was operated at 200 keV with electron beam current ~40 pA, a convergence semi-angle of 10 mrad, and a camera length of 91 mm. A total of 129 projections were collected with a tilt range from -58° and +70° in 1° increments. After background subtraction, the projections were aligned along the tilt axis direction by cross-correlation and along the perpendicular direction using the center-of-mass method[9]. Reconstructions were performed with GENFIRE and SIRT. The SIRT reconstruction was computed using Tomo3D[52]. Figures 5a and b show the 13.6-nm-thick central slice of the GENFIRE and SIRT reconstruction of a fragment of the sample, revealing a complex 3D porous structure. Along the 0° direction, both GENFIRE and SIRT produce good reconstructions, although fine features are better resolved by GENFIRE (Figs. 5a and b). However, in the missing wedge direction, GENFIRE exhibits significant improvement over SIRT with sharper boundaries and more distinctive 3D pore structures (Figs. 5c and d). Figures 5e and f show isosurface renderings of the reconstructions, where elongation artifacts due to the missing wedge are clearly visible in the SIRT reconstruction, but are reduced by GENFIRE.

**GENFIRE reconstruction of a frozen hydrated cell**. GENFIRE was also used to reconstruct the 3D structure of a frozen-hydrated marine cyanobacterium in a late stage of infection by cyanophages[53]. A tilt series of 42 projections ranging from -58° to +65° were acquired on a JEM2200FS electron microscope equipped with a Zernike phase plate and recorded on a 4k x 4k Gatan CCD[53]. The projections were binned by 4x4 pixels, resulting in images with approximately $1.8 \times 1.8$ nm$^2$ per pixel. The background was carefully removed from each projection based on the average value in a flat region outside of the cell. A marine cyanobacterium was then cropped out from the surrounding regions by smoothing and thresholding each projection to produce a soft-edged mask. Finally, each projection was aligned and normalized to have the same total sum as the integrated density should be conserved. The tilt series was separately reconstructed with GENFIRE and FBP (Fig. 6). The GENFIRE reconstruction was performed for 100 iterations with a loose cubic support, while the FBP reconstruction was computed using IMOD[54]. Several low-contrast features are visible in the GENFIRE reconstruction that are difficult, if not impossible, to identify with FBP. Of particular interest in this dataset was the interactions between the marine cyanobacterium and cyanophages. Fig. 6 shows a slice through the reconstructed volumes capturing the penetration of a cyanophage into the cell membrane during the infection process. This interaction has caused a local depression in the cell membrane, and the shown cross section passes through this depression as well as the viral capsid and appendage (Figs. 6c-j). Based on this geometry the cell membrane should be visible on both sides of the interaction, similar to taking a horizontal cross-section through a U-shape (Figs. 6c and d). Although the top side of the membrane is visible in both reconstructions (magenta arrows), the bottom side is only visible in the GENFIRE reconstruction (yellow arrow). Figures 6i and j show isosurface renderings of the penetration of the cyanophage into the cell membrane, where GENFIRE exhibits higher contrast, less peripheral noise, more easily detectable cell boundaries than FBP.

**Discussions**

In this article, we present the mathematical implementation of GENFIRE for 3D reconstruction from a limited number of projections with a missing wedge. Both numerical simulation and experimental results of materials science and biological specimens indicate that GENFIRE produces superior 3D reconstruction to several other tomographic algorithms. As a Fourier-based iterative method, GENFIRE first computes a small fraction of Cartesian grid points with high precision from 2D projections using Fourier gridding and oversampling. It then iterates between real and reciprocal space using the FFT and its inversion. Positivity and support are enforced in real space, while the grid points calculated from the measured data are applied in reciprocal space. As the Fourier data, positivity and support are all convex constraint sets, GENFIRE belongs to the method of projections onto convex sets, whose convergence has been mathematically proven[55,56]. This allows GENFIRE to search for a global solution that is concurrently consistent with the measured data and physical constraints. One of the unique features of Fourier-based iterative algorithms such as GENFIRE is that any changes in real space globally affect all the points in reciprocal space and vice versa. This global correlation between real and reciprocal space makes GENFIRE robust to the missing data and missing wedge. In contrast, ART, SART and SIRT perform all the iterations in real space through local interpolation. When there is a missing wedge, the local interpolation in that region becomes less accurate. This explains why GENFIRE achieves better 3D reconstructions than several other tomographic algorithms. Furthermore, compared to EST that is only applicable to single tilt axis data, GENFIRE can not only work with any tomographic geometry, but also perfroms faster due to the use of the FFT and its inversion for iteration.

Another class of tomographic reconstruction methods based on compressed sensing is presently under rapid development[57,58]. Compressed sensing assumes that a physically meaningful structure is usually sparse in some domain. If the sparse domain can be found, the 3D structure can in principle be reconstructed from a small number of 2D projections. Compressed sensing tomography typically incorpoartes mathematical regularization such as total variation minimization[59], which requires manual tuning of parameters[60]. This is acceptable in certain applications, where the scope of reconstruction targets is limited enough to permit a specialized set of parameters. However, for general tomographic reconstructions, it is not straightforward to optimize these parameters, especially with the presecense of missing data and noise. For example, it would be very challanging, if not impossible, for compressed sensing tomography to reconstruct the 3D distribution of point defects in a crystalline specimen. Conversely, GENFIRE uses very general physical constraints and requires minimum manual tuning of parameters. It has recently been used to determine crystal defects such as grain boundaries, chemical order/disorder, anti-phase boundaries and point defects with unprecendent 3D detail[10,13]. Furthermore, GENFIRE can be easily adapted to incorporate mathematical regularization to reconstruct 3D sparse objects from a small number of projections. Looking forward, we expect GENFIRE can be applied to a plethora of imaging modalities to address a wide range of scientific problems.

**References**

1. Miao, J., Ercius, P. & Billinge, S. J. L. Atomic electron tomography: 3D structures without crystals. *Science* **353,** (2016).
2. Frank, J. *Three-Dimensional Electron Microscopy of Macromolecular Assemblies: Visualization of Biological Molecules in Their Native State*.
3. Kak, A. C. & Slaney, M. *Principles of computerized tomographic imaging*. (Society for Industrial and Applied Mathematics, 2001). doi:10.1137/1.9780898719277
4. Herman, G. T. & Herman, G. T. *Fundamentals of computerized tomography: image reconstruction from projections*. (Springer, 2009).
5. Momose, A., Takeda, T., Itai, Y. & Hirano, K. Phase-contrast X-ray computed tomography for observing biological soft tissues. *Nat. Med.* **2,** 473–475 (1996).
6. Pfeiffer, F., Weitkamp, T., Bunk, O. & David, C. Phase retrieval and differential phase-contrast imaging with low-brilliance X-ray sources. *Nat. Phys.* **2,** 258–261 (2006).
7. Fernandez-Leiro, R. & Scheres, S. H. W. Unravelling biological macromolecules with cryo-electron microscopy. *Nature* **537,** 339–346 (2016).
8. Scott, M. C. *et al.* Electron tomography at 2.4-angstrom resolution. *Nature* **483,** 444–447 (2012).

9. Chen, C.-C. *et al.* Three-dimensional imaging of dislocations in a nanoparticle at atomic resolution. *Nature* **496,** 74–77 (2013).
10. Xu, R. *et al.* Three-dimensional coordinates of individual atoms in materials revealed by electron tomography. *Nat. Mater.* **14,** 1099–1103 (2015).
11. Goris, B. *et al.* Measuring lattice strain in three dimensions through electron microscopy. *Nano Lett.* **15,** 6996–7001 (2015).
12. Haberfehlner, G. *et al.* Formation of bimetallic clusters in superfluid helium nanodroplets analysed by atomic resolution electron tomography. *Nat. Commun.* **6,** 8779 (2015).
13. Yang, Y. *et al.* Deciphering chemical order/disorder and material properties at the single-atom level. *Nature* **542,** 75–79 (2017).
14. Cheng, Y. Single-particle cryo-EM at crystallographic resolution. *Cell* **161,** 450–457 (2015).
15. Nogales, E. The development of cryo-EM into a mainstream structural biology technique. *Nat Methods* **13,** (2016).
16. Bartesaghi, A., Matthies, D., Banerjee, S., Merk, A. & Subramaniam, S. Structure of β-galactosidase at 3.2-Å resolution obtained by cryo-electron microscopy. *Proc. Natl. Acad. Sci. U. S. A.* **111,** 11709–11714 (2014).
17. Lučić, V., Förster, F. & Baumeister, W. Structural studies by electron tomography: from cells to molecules. *Annu. Rev. Biochem.* **74,** 833–865 (2005).
18. Oikonomou, C. M. & Jensen, G. J. A new view into prokaryotic cell biology from electron cryotomography. *Nat. Rev. Microbiol.* **14,** 205–220 (2016).
19. Lee, E. *et al.* Radiation dose reduction and image enhancement in biological imaging through equally-sloped tomography. *J. Struct. Biol.* **164,** 221–227 (2008).
20. Benjamin P Fahimian and Yu Mao and Peter Cloetens and Jianwei Miao. Low-dose x-ray phase-contrast and absorption CT using equally sloped tomography. *Phys. Med. Biol.* **55,** 5383 (2010).
21. Larabell, C. A. & Nugent, K. A. Imaging cellular architecture with X-rays. *Curr. Opin. Struct. Biol.* **20,** 623–631 (2010).
22. Zhao, Y. *et al.* High-resolution, low-dose phase contrast X-ray tomography for 3D diagnosis of human breast cancers. *Proc. Natl. Acad. Sci.* **109,** 18290–18294 (2012).
23. Gibbs, J. W. *et al.* The three-dimensional morphology of growing dendrites. *Sci. Rep.* **5,** (2015).
24. Meirer, F. *et al.* Three-dimensional imaging of chemical phase transformations at the nanoscale with full-field transmission X-ray microscopy. *J. Synchrotron Radiat.* **18,** 773–781 (2011).
25. Krenkel, M. *et al.* Phase-contrast zoom tomography reveals precise locations of macrophages in mouse lungs. *Sci. Rep.* **5,** (2015).
26. Miao, J., Charalambous, P., Kirz, J. & Sayre, D. Extending the methodology of X-ray crystallography to allow imaging of micrometre-sized non-crystalline specimens. *Nature* **400,** 342–344 (1999).
27. Miao, J. *et al.* Three-dimensional GaN-Ga2O3 core shell structure revealed by X-ray diffraction microscopy. *Phys. Rev. Lett.* **97,** (2006).
28. Nishino, Y., Takahashi, Y., Imamoto, N., Ishikawa, T. & Maeshima, K. Three-dimensional visualization of a human chromosome using coherent X-ray diffraction. *Phys. Rev. Lett.* **102,** 18101 (2009).
29. Jiang, H. *et al.* Quantitative 3D imaging of whole, unstained cells by using X-ray diffraction microscopy. *Proc. Natl. Acad. Sci. U. S. A.* **107,** 11234–11239 (2010).
30. Dierolf, M. *et al.* Ptychographic X-ray computed tomography at the nanoscale. *Nature* **467,** 436–439 (2010).
31. Jiang, H. *et al.* Three-dimensional coherent X-ray diffraction imaging of molten iron in mantle olivine at nanoscale resolution. *Phys. Rev. Lett.* **110,** 205501 (2013).
32. Miao, J., Ishikawa, T., Robinson, I. K. & Murnane, M. M. Beyond crystallography: Diffractive imaging using coherent x-ray light sources. *Science* **348,** 530–535 (2015).
33. Holler, M. *et al.* High-resolution non-destructive three-dimensional imaging of integrated circuits. *Nature* **543,** 402–406 (2017).

34. Gordon, R., Bender, R. & Herman, G. T. Algebraic Reconstruction Techniques (ART) for three-dimensional electron microscopy and X-ray photography. *J. Theor. Biol.* **29,** 471–481 (1970).
35. Andersen, A. Simultaneous Algebraic Reconstruction Technique (SART): A superior implementation of the ART algorithm. *Ultrason. Imaging* **6,** 81–94 (1984).
36. Gilbert, P. Iterative methods for the three-dimensional reconstruction of an object from projections. *J. Theor. Biol.* **36,** 105–117 (1972).
37. Miao, J., Förster, F. & Levi, O. Equally sloped tomography with oversampling reconstruction. *Phys. Rev. B* **72,** 52103 (2005).
38. O'Connor, Y. Z. & Fessler, J. A. Fourier-based forward and back-projectors in iterative fan-beam tomographic image reconstruction. *IEEE Trans. Med. Imaging* **25,** 582–589 (2006).
39. Fahimian, B. P. *et al.* Radiation dose reduction in medical x-ray CT via Fourier-based iterative reconstruction: Dose reduction in CT via Fourier-based iterative reconstruction. *Med. Phys.* **40,** 31914 (2013).
40. Miao, J., Sayre, D. & Chapman, H. N. Phase retrieval from the magnitude of the Fourier transforms of nonperiodic objects. *J. Opt. Soc. Am. A* **15,** 1662 (1998).
41. Miao, J. & Sayre, D. On possible extensions of X-ray crystallography through diffraction-pattern oversampling. *Acta Crystallogr. A* **56 (Pt 6),** 596–605 (2000).
42. Shepard, D. A two-dimensional interpolation function for irregularly-spaced data. in 517–524 (ACM Press, 1968). doi:10.1145/800186.810616
43. Franke, R. Scattered data interpolation: tests of some methods. *Math. Comput.* **38,** 181–181 (1982).
44. Brünger, A. T. Free R value: cross-validation in crystallography. *Methods Enzymol.* **277,** 366–396 (1997).
45. Rossmann, M. G. Ab initio phase determination and phase extension using non-crystallographic symmetry. *Curr. Opin. Struct. Biol.* **5,** 650–655 (1995).
46. Raines, K. S. *et al.* Three-dimensional structure determination from a single view. *Nature* **463,** 214–217 (2010).
47. Kirkland, E. J. *Advanced computing in electron microscopy*. (Springer, 2010).
48. Dabov, K., Foi, A., Katkovnik, V. & Egiazarian, K. Image Denoising by Sparse 3-D Transform-Domain Collaborative Filtering. *IEEE Trans. Image Process.* **16,** 2080–2095 (2007).
49. Huynh, D. Q. Metrics for 3D rotations: comparison and analysis. *J. Math. Imaging Vis.* **35,** 155–164 (2009).
50. Lemberg, J. A. & Ritchie, R. O. Mo-Si-B alloys for ultrahigh-temperature structural applications. *Adv. Mater.* **24,** 3445–3480 (2012).
51. Rioult, F. A., Imhoff, S. D., Sakidja, R. & Perepezko, J. H. Transient oxidation of Mo–Si–B alloys: Effect of the microstructure size scale. *Acta Mater.* **57,** 4600–4613 (2009).
52. Agulleiro, J.-I. & Fernandez, J.-J. Tomo3D 2.0 – exploitation of advanced vector extensions (AVX) for 3D reconstruction. *J. Struct. Biol.* **189,** 147–152 (2015).
53. Dai, W. *et al.* Visualizing virus assembly intermediates inside marine cyanobacteria. *Nature* **502,** 707–710 (2013).
54. Mastronarde, D. N. Dual-axis tomography: an approach with alignment methods that preserve resolution. *J. Struct. Biol.* **120,** 343–352 (1997).
55. Youla, D. C. & Webb, H. Image restoration by the method of convex projections. *IEEE Trans. Med. Imaging* **1,** 81–94 (1982).
56. Sezan, M. I. An overview of convex projections theory and its application to image recovery problems. *Ultramicroscopy* **40,** 55–67 (1992).
57. Candes, E. J., Romberg, J. & Tao, T. Robust uncertainty principles: exact signal reconstruction from highly incomplete frequency information. *IEEE Trans. Inf. Theory* **52,** 489–509 (2006).
58. Donoho, D. L. Compressed sensing. *IEEE Trans. Inf. Theory* **52,** 1289–1306 (2006).
59. Rudin, L. I., Osher, S. & Fatemi, E. Nonlinear total variation based noise removal algorithms. *Phys. Nonlinear Phenom.* **60,** 259–268 (1992).

60. Leary, R., Saghi, Z., Midgley, P. A. & Holland, D. J. Compressed sensing electron tomography. *Ultramicroscopy* **131,** 70–91 (2013).

**Acknowledgements**

We thank M. Taylor and J. Perepezko for the preparation of the $Mo_3Si$ alloy sample. This work was supported by STROBE: A National Science Foundation Science & Technology Center under Grant No. DMR 1548924, the Office of Basic Energy Sciences of the US DOE (DE-SC0010378), the NSF DMREF program (DMR-1437263), and the DARPA PULSE program through a grant from AMRDEC. Work at the Molecular Foundry was supported by the Office of Science, Office of Basic Energy Sciences, of the U.S. Department of Energy under Contract No. DE-AC02-05CH11231.

**Software availability.** The GENFIRE software package with a graphical user interface will be freely available at www.physics.ucla.edu/research/ imaging/ GENFIRE.

**Figure legends**

**Figure 1**. The GENFIRE algorithm. GENFIRE first computes oversampled Fourier slices from a tilt series of 2D projections. The oversampled Fourier slices are used to accurately calculate a small fraction of points on a 3D Cartesian grid based on gridding interpolation. The algorithm then iterates between real and reciprocal space. The support and positivity constraints are enforced in real space, while the small fraction grid points corresponding to the measured data are enforced in reciprocal space. Error metrics are used to monitor the convergence of the iterative process. After several hundred iterations, the algorithm converges to a 3D structure that is concurrently consistent with the measured data in reciprocal space and the physical constraints in real space.

**Figure 2**. Numerical simulations on the 3D reconstruction of a biological vesicle from 71 noisy projections using GENFIRE, EST, FBP and SIRT. **a-c**, Three 10-voxel-thick central slices of the vesicle model in the XY, ZX and ZY planes, respectively. The corresponding three reconstructed slices with GENFIRE (**d-f**), EST (**g-i**), FBP (**j-l**), and SIRT (**m-o**), where the missing wedge axis is along the z-axis. **p**, The FSC between the reconstructions and the model, showing that GENFIRE produces a more faithful reconstruction than other algorithms at all spatial frequencies.

**Figure 3**. Numerical simulations on atomic electron tomography. 1.2-Å-thick central slices of a $L1_0$ phase FePt nanoparticle in the XY and ZX planes, reconstructed from 27 noisy multislice STEM projections with GENFIRE (**a, e**), EST (**b, f**), SIRT (**c, g**), and FBP (**d, h**), where the z-axis is the missing wedge direction. The red arrow indicates a Pt atom and the white arrow an Fe atom. A central slice in the ZX plane after applying the Fourier transform to the 3D reconstruction obtained by GENFIRE (**i**), EST (**j**), SIRT (**k**), and FBP (**l**), showing recovery of the Bragg peaks in the missing wedge direction for GENFIRE, EST and SIRT (magenta arrows). Artifacts due to missing wedge effects such as "ghost atoms" are visible in SIRT and FBP (**c, d, g,** and **h**), but are not present in EST and GENFIRE (**a**, **b**, **e** and **f**).

**Figure 4**. Angular refinement simulations for the GENFIRE reconstruction of the 27 multislice STEM projections used in Fig. 3. **a**, The angular error between correct projection angles and the initial misaligned angles (black dots) and the refined ones after 5 refinement iterations (blue crosses), improving an average angular error from 2.1° to 1.3°. **b**, **c**, 1.2-Å-thick central slices before and after angular refinement, showing some Fe atoms in the lower left region are better resolved and the boundary of the nanoparticle is also better defined.

**Figure 5**. Comparison of GENFIRE and SIRT reconstructions of a fragment of porous $Mo_3Si$ alloy, annealed at 1100° C. **a**, **b**, 13.6-nm-thick central slices along 0° direction reconstructed by GENFIRE and SIRT, respectively, where fine features are better resolved in the GENFIRE reconstruction. **c**, **d**, 13.6-nm-thick central slices of the GENFIRE and SIRT reconstructions along the missing wedge direction, where GENFIRE shows significant improvement over SIRT with sharper boundaries and more distinctive 3D pore structures. **e**, **f**, Isosurface renderings of GENFIRE and SIRT reconstructions, where elongation artifacts due to the missing wedge are visible in the SIRT reconstruction, but are reduced by GENFIRE.

**Figure 6**. 3D structure of a frozen-hydrated marine cyanobacterium, capturing the penetration of a cyanophage into the cell membrane. **a**, **b**, 5.4-nm-thick slices of the cell in the XY plane reconstructed by GENFIRE and FBP, respectively. Magnified views of the penetration of a cyanophage for the GENFIRE and FBP reconstructions in the XY (**c, d**), XZ (**e, f**), and ZY (**g, h**) planes, respectively. The top side of the membrane is visible in both reconstructions (magenta arrows), but the bottom side is only visible with GENFIRE (yellow arrow). **i**, **j**, Isosurface renderings of the penetration of the cyanophage to the cell membrane. Overall, GENFIRE exhibits higher contrast, less peripheral noise, more easily detectable cell boundaries than FBP.

**Figures**

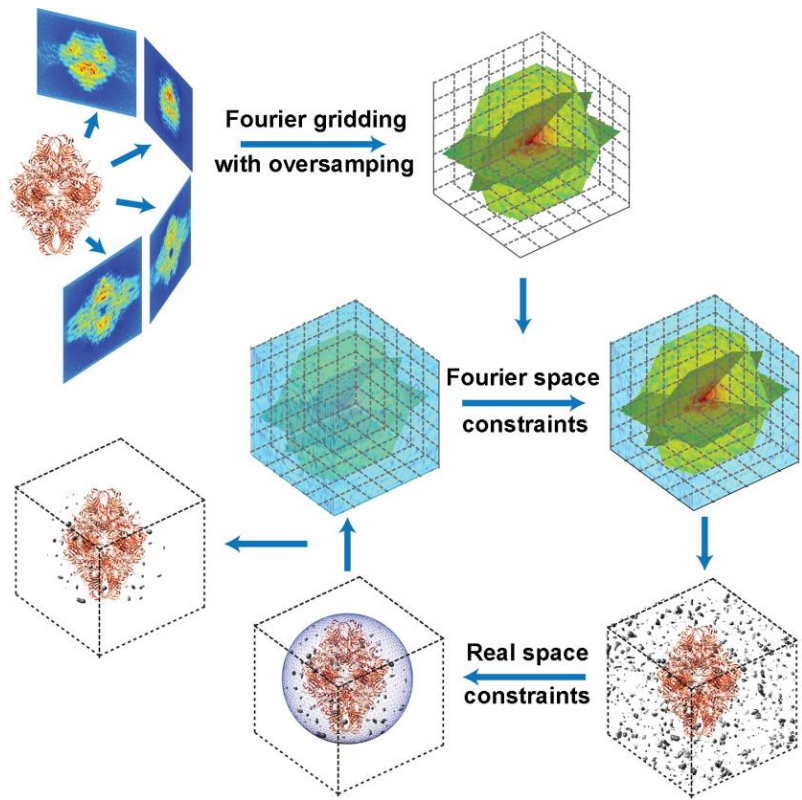

**Fig. 1**

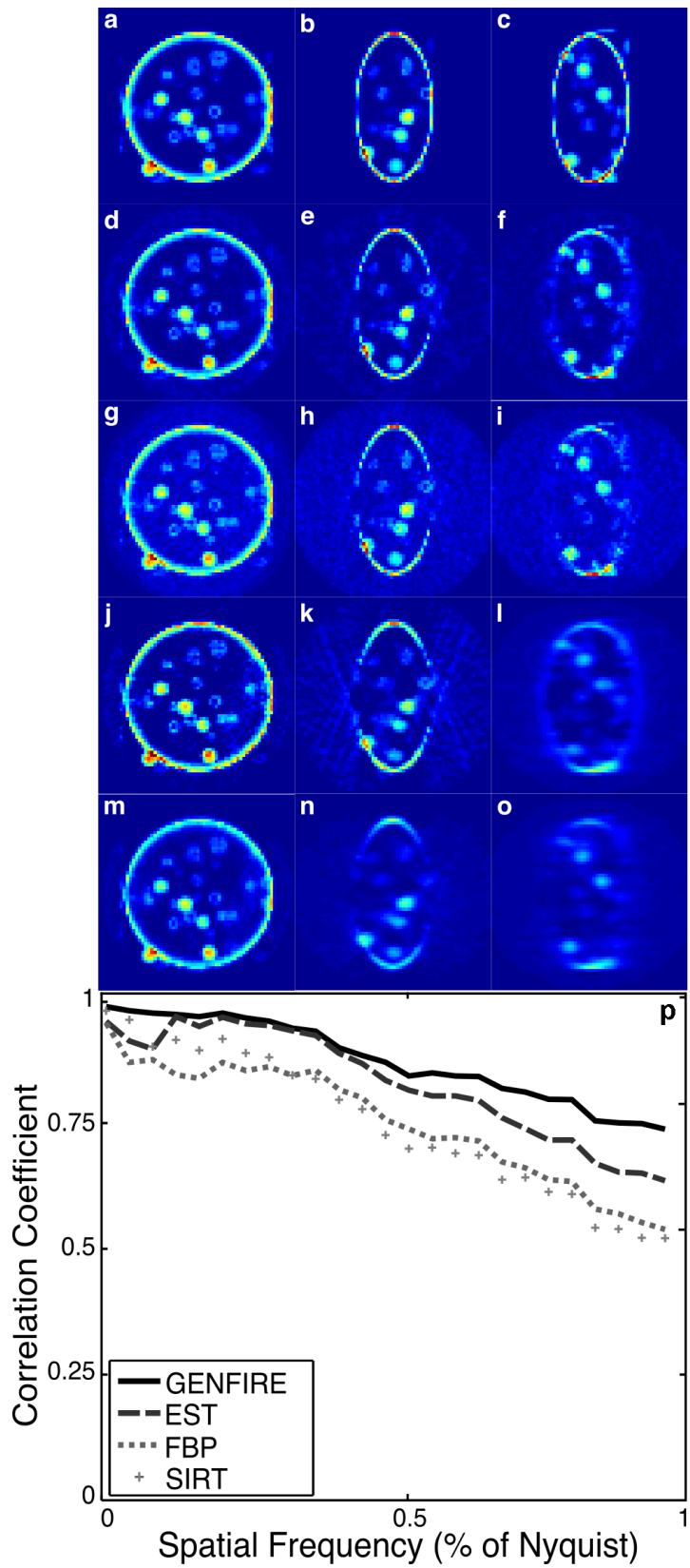

Fig. 2

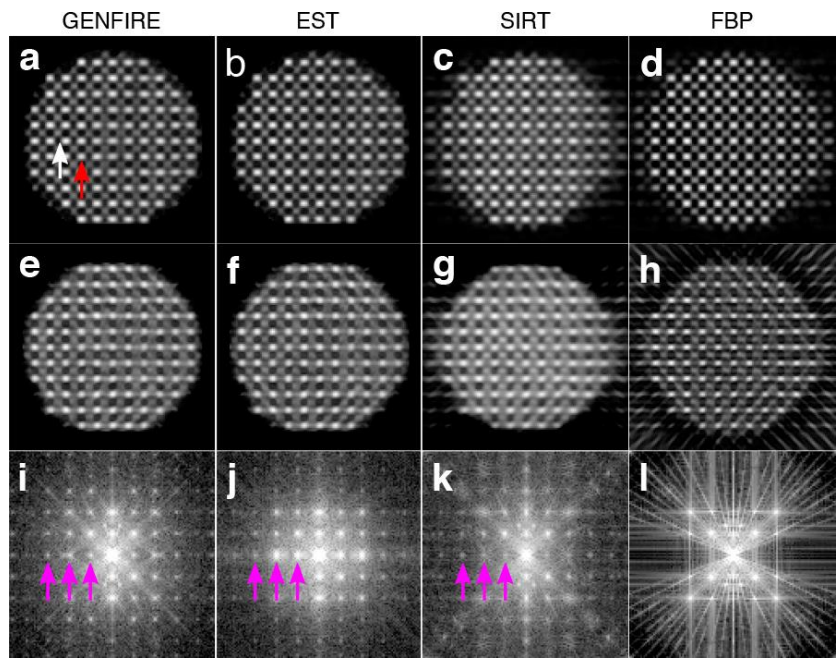

Fig. 3

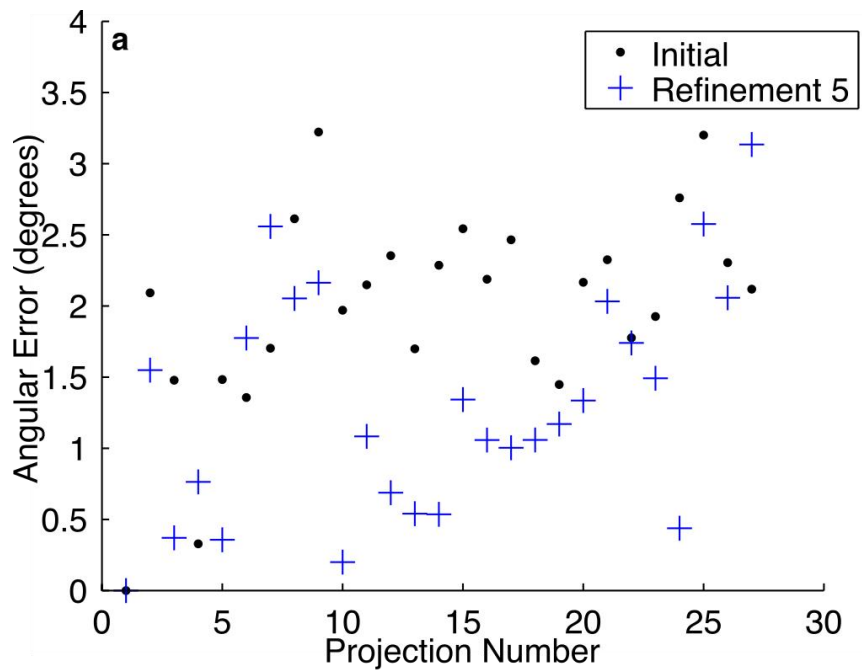
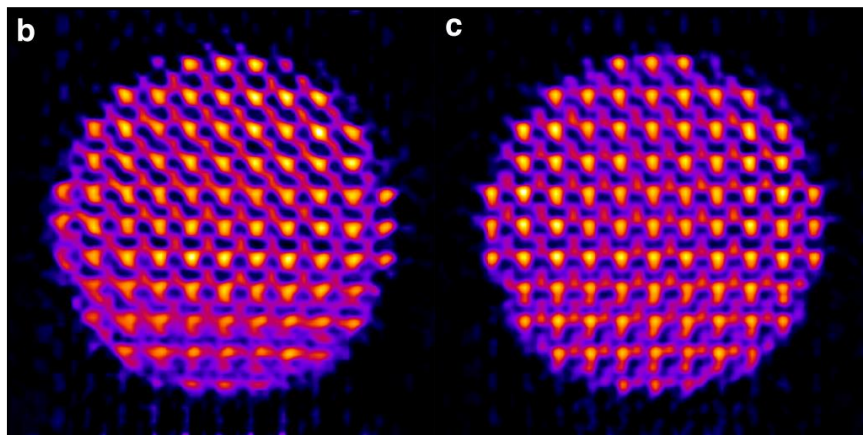

**Fig. 4**

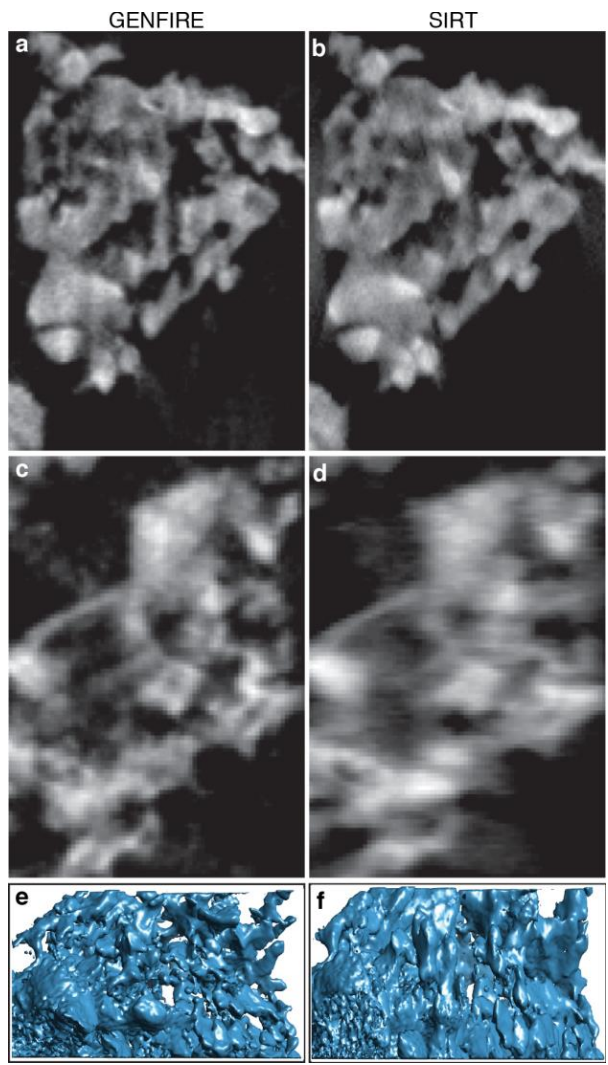

**Fig. 5**

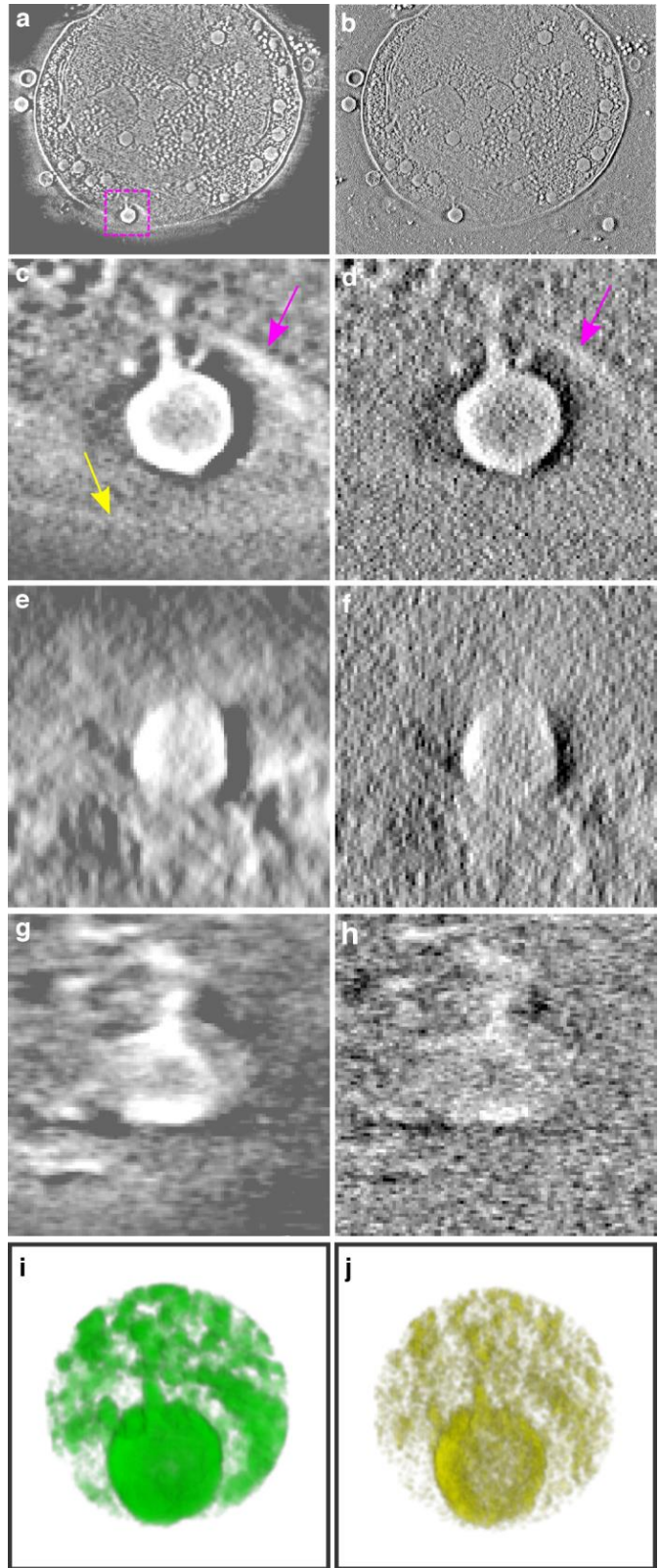

**Fig. 6**